\documentclass[twoside]{article}
\usepackage{fleqn,espcrc2}
\usepackage{amsmath}
\usepackage{graphicx} % if you want to include PostScript figures

\newcommand{\fcaption}{%
\vspace*{-1.0cm}
\caption%
}

\title{ New Numerical Methods for Quantum Field Theories on the
Continuum\thanks{ Presented by P. Emirda\u{g} at \it{Lattice '99}, Pisa
(Italy).}}

\author{P. Emirda\u{g}$^{\mathrm a}$, R. Easther$^{\mathrm a}$, G. S.
Guralnik$^{\mathrm a}$, S. C. Hahn%
\address{Department of Physics, Brown University, Providence, RI, USA 02912--1843}}

\begin{document}
\begin{abstract}
The Source Galerkin Method is a new numerical technique that is being
developed
 to solve Quantum Field Theories on the
continuum. It is not based on Monte Carlo techniques
and has a measure to evaluate relative errors. It
promises to increase  the  accuracy and speed of calculations,
and  takes full advantage of symmetries of the theory. %[RE -reworded]
The application of this method to the non-linear $\sigma$ model is outlined.
\end{abstract}

\maketitle

\section{MOTIVATION}
Conventionally, in numerical approaches to field theories,
integrals for generating functionals are evaluated on lattices using
Monte Carlo methods. These are quite successful but there are still
constraints due to lack of computer power and algorithmic
limitations.  These limitations become important for theories 
with significant contributions from fermionic interactions
beyond the quenched approximation, and for actions that are not positive
definite. Source Galerkin is an alternative computational method
that avoids these particular problems \cite{SGL94,GL2}. It works both on
the lattice and the continuum and  handles fermions as easily as bosons
\cite{HG1,HG2}. Computations require significantly less time than Monte
Carlo based methods. The basic idea of the method is to find solutions
to the functional differential equations for a field theory
in the presence of sources. The solutions are generating functionals of
the theory.

\subsection{Overview of the Method}
To analyze a theory,  we start with a Lagrangian in the
presence of  external sources.  The source functional differential
equations satisfied by $Z$ are generated in the usual way. The Source
Galerkin technique consists of a systematic iterative way to solve these
equations to increasing accuracy.  We proceed by introducing an ansatz
which is an expansion over the external sources of the theory,
\begin{equation*}
Z^{\ast}= \exp  \big[\int j_x G_{xy}j_y +\int j_{\omega}j_x H_{\omega xyz}j_y j_z + \cdots \big]
\end{equation*}
$Z^*$ does not satisfy the source differential equations of the field
theory exactly, but we can require that specific functional
projections of these equations are satisfied. These projections are
formed by multiplying by test functions in sources and integrating
this product in source space using a fixed measure. Introducing the
appropriate number of test functions, we can solve for the parameters
of the resulting expressions.  This method causes the error in assumed
solutions to approach zero in the mean and so the ans\"atz satisfies the
equations approximately. The details of this will be made clear in what
follows. This approach belongs to the general class of techniques called
Galerkin methods \cite{FLE}. 
%[RE - put citation to Galerkin methods reference here]

\section{APPLICATIONS}
The method has been applied to various field theories on the lattice
and, more recently, on the continuum $\lambda\phi^{4}$, Gross-Neveu,
and other two dimensional fermionic models have been studied. Here, we
discuss an application of the method to the $O(3)$ Nonlinear $\sigma$
Model with $d=2$. Since this model is asymptotically free, it serves
as a toy model for our initial approach to Non-Abelian gauge theories.

\subsection{$O(3)$ Nonlinear $\sigma$ Model ($d=2$)}
\begin{figure}
\includegraphics[width=\linewidth,angle=270]{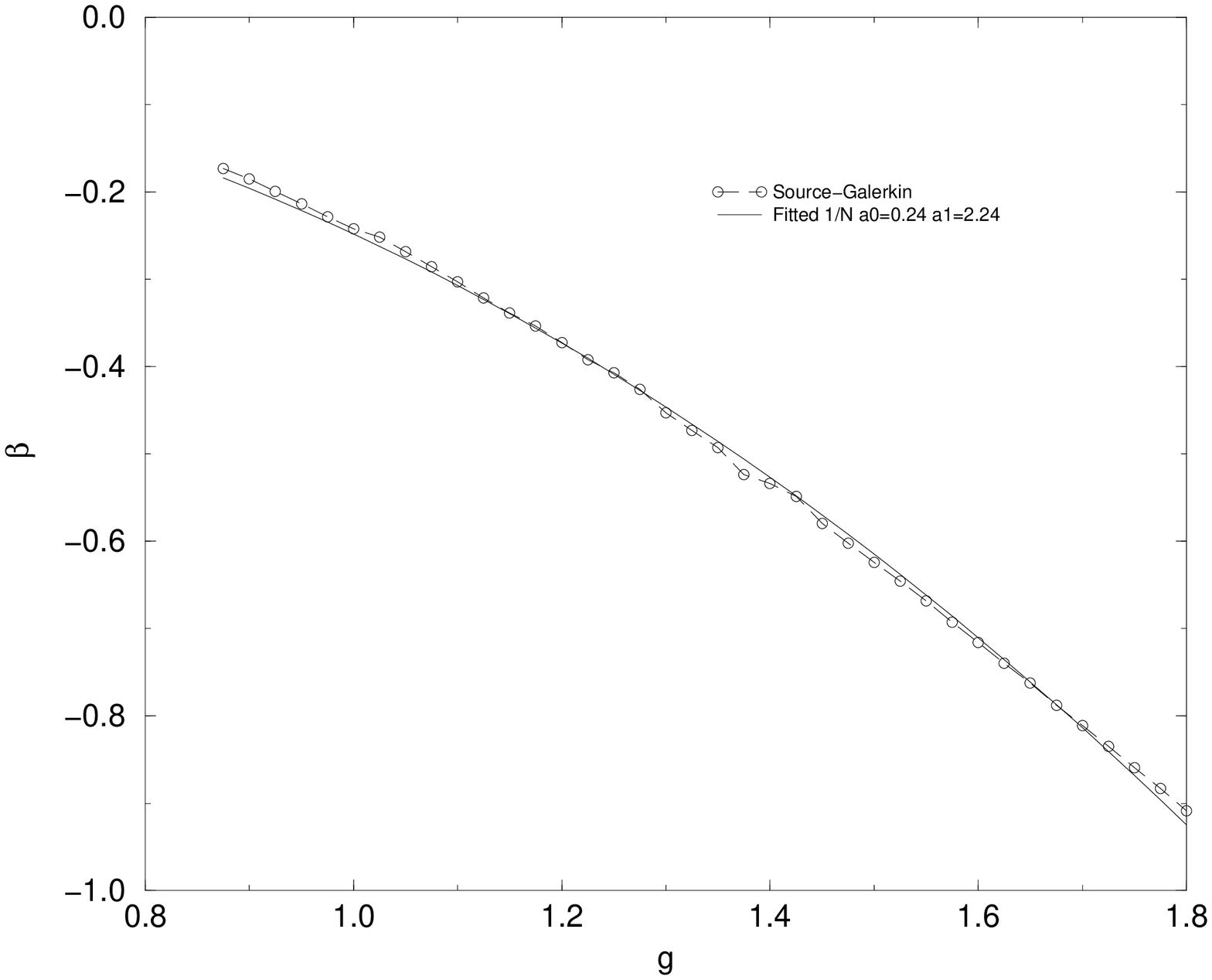}%
\fcaption{$\beta$ function vs. coupling, compared to $1/N$ result.}
\label{beta}
\end{figure}

This model consists of three real fields coupled to each other.
The Lagrangian is:
\begin{multline*}
 L = \frac{1}{2} \partial_
{\mu}\phi^{a}(x)\partial^{\mu}\phi^{a}(x)+
 j^{a}(x)\phi^{a}(x) \\
  +  \frac{1}{2}  \chi(x)
\left( \phi^{a}(x)\phi^{a}(x)-\frac{1}{g^2} \right)-\chi(x)S
(x)
\end{multline*}
$\phi$ and $\chi$ are the canonical and Lagrange multiplier fields
respectively. $j$ and $S$ are the external sources for these fields
while $g$ is the coupling constant of the theory.  The Schwinger--Dyson
equations are
\begin{gather*}
 \partial^{2} \frac{\delta Z[j,S]}
{\delta j^{a}(x)}- \frac{\delta^{2}Z[j,S]}{\delta j^{a}(x)
 \delta S(x)}
- j^{a}(x)Z[j,S]=0,\\
 \frac{\delta^{2} Z[j,S] }{ \delta j^{a}(x) \delta j^{a}(x)}
- \frac{1}{g^2}Z[j,S]- 2 S(x)Z[j,S]=0.
\end{gather*}
These are the functional differential equations to which we apply the
Galerkin procedure. The simplest ans\"atz is
\begin{equation*}
Z = \exp\left[\frac{1}
{2}\int j^{a}(x)G^{ab}(x,y)j^{b}(y) +
\chi_{0}\int S(x)\right].
\end{equation*}
The inner product that will be used for projecting residuals on the test
functions is
\begin{equation*}
(A[j],B[j])_{j}= \int [dj] e^{-j^{2}/\epsilon^{2}}A[j]B[j] .
\end{equation*}
The subscript denotes the $j$ space. 
 $\epsilon$ weights the inner product such that higher order
terms contribute less. For this model, the projections are
\begin{gather*}
(1,(j^{b}(x^{\prime}),R_{a})_{j})_{S} = 0,\\
(1,(1,R_{c})_{j})_{S} = 0.
\end{gather*}
$R_{a}$ and $R_{c}$ are the residuals for the equations of motion and
constraint respectively.  These equations, are solved for Green's functions
and Lagrange multiplier, $\chi_{0}$, and are defined in source and
coordinate space. The space-time dependence of the Green's functions is
approximated by truncating a Sinc expansion,
\begin{equation*}
G(x)  = \sum_{k=-N}^{N}G(kh)S(k,h)(x).
\end{equation*}
where the Sinc function is defined as
\begin{equation*}
S(k,h)(x)  =   \frac{ \sin{\pi(x-kh)/h}}{ \pi(x-kh)/h}.
\end{equation*}
%This is a good approximation for functions that fall off
%algebraically or exponentially at $x\rightarrow \infty$.
% sch: this sentence could be strengthened or dropped
Sinc functions have many algebraic properties
that make them very useful to work with \cite{STE}.
%In two dimensions;
%\begin{equation*}
%G(x,y)=\sum_{k,l}G(kh,lh)S(k,h)(x)S(l,h)(y)
%\end{equation*}
%[RE- Define "N" is the equations below, since it refers (I think) to the
%truncation of the Sinc functions, not to $\mathcal{O}(N)$]
% sch: agree
While this is a very direct approach, this interpolative  approximation
requires that every extra dimension introduces another sum to the problem.
 Using this approximation $\mathcal{O}(N^{d})$ numbers are required to store a
two point function. 
%This is manageable for simplest ans\"atz but
The memory
limitations become a concern when  $\mathcal{O}(N^{6d})$ numbers are to be stored for
the Jacobian with four-point functions. This leads us to seek an alternative
method which uses general information about the spectral structure of
field theories. Any exact two point function can be represented as a sum
over free two point functions using spectral representations. Consequently,
we can shift our basis for numerical solutions to regulated free  Euclidean
Green's functions:
\begin{equation*}
G(x)=\frac{1}{(2 \pi)^d} \int dp\,\frac{e^{-p^2/\Lambda^2+ipx}}{p^2+m^2}
\end{equation*}
A term in the spectral representations has the Sinc expansion:
\begin{align*}
G(x) &\approx G_{0}\sum\limits_{k=-N}^{N} c(k) \exp \left[-\frac{x^{2}}{4(e^{kh}
+1/ \Lambda^{2})} \right] \\ 
c(k) &= \frac{1}{e^{kh}}\left[\frac{\pi}{e^{kh}+1/\Lambda^{2}}
\right]^{d/2}\exp\left[ -e^{kh}m^{*^2} \right]
\end{align*}
The Galerkin equations are solved approximately using only one term of the spectral expansion with the parameters  $m^{*^2} $ and $ G_{0}$.
%In this approximation space-time dimensionality of the  problem does not
%introduce an extra sum. 
The $\beta$ function (Figure.~\ref{beta}) is
calculated and 
% using the Callan-Symanzik equation.
%\begin{equation*}
%\left[ \Lambda\frac{\partial}{\partial
%\Lambda}+\beta(g)\frac{\partial}{\partial g} \right]m(\Lambda,g)=0
%\end{equation*}
 compared to  $1/N$ results.
\begin{figure}
\includegraphics[width=\linewidth]{phi4-fourth-graphs.0}%
\fcaption{Comparison of improved ans\"atz with lowest order and exact answer 
for $\lambda\phi^{4}$, $d=1$}
\label{imp}
\end{figure}
The ans\"atz is improved, 
\begin{multline*}
Z^{\ast} = Z^{\ast}_{s} \exp \left[ \frac{1}{2} \int S(x)D(x-y)S(y)\right.\\
+ \left.\frac{1}{2} \int j^a(x)j^b(y)F^{ab}(x-z,y-z)S(z) \right].
\end{multline*}
$Z_{s}$ is the simplest ans\"atz. $F^{ab}$ and $D$ are the vertex term
and the two-point function for the $\chi$ field respectively. Results get close to exact value dramatically with improved ans\"atz. 
%Moving
%towards the exact value drastically by improving the ans\"atz is an
%indication of the success of the method.  
We have shown this in the $\lambda \phi^{4} $ model 
 by introducing a four-point function 
(Figure.~\ref{imp}).
%\begin{figure}[!hb]
%\includegraphics[width=\linewidth]{phi4-fourth-graphs.0}%
%\fcaption{Comparison of improved ans\"atz with lowest order and exact answer
%for $\lambda\phi^{4}$, $d=1$}
%\label{imp}
%\end{figure}

\section{CONCLUSIONS}
The Source Galerkin method promises to extend the range of
field theoretic problems that can be solved with current computers.
The low order results of the method applied to the nonlinear $\sigma$
model show asymptotic freedom and are close to the large-$N$ results.
The contributions from higher order corrections are being studied.
% and
% are essential to confirm the validity of the work to date.
Understanding this model is of primary importance, as it is a step to
supersymmetric and matrix models. One of the major advantages of our
method lies in its approach to fermionic fields. Fermions are treated
identically to bosons except for the anti-commutivity of the
associated sources. Consequently, expansions involving fermions and
bosons are  symmetric in both types of fields and there is
no analogue to the fermion determinant problem. The preliminary
results of fermionic theories are very promising \cite{GL1}.  Source
Galerkin takes advantage of the symmetries of a theory.
% when setting up
%the ans\"atz for the generating functional.  
The calculations are done
in a continuum formulation, thus avoiding lattice-specific
problems. While studying various ways of approximating Green's
functions an interesting application has been investigated.  Using
Sinc-based spectral representations, we can calculate Feynman diagrams
in a
%regulated [RE - there is no need to say "regulated" here,
%since it implies that we can't renormalize] 
perturbation theory. Complicated diagrams can be evaluated much
faster than with  Monte-Carlo methods, and to higher accuracies \cite{FEYN}.

\section*{ACKNOWLEDGMENTS}
Computational work in support of this research was performed at the Theoretical Physics Computing Facility at Brown University and National Energy Research Scientific Computing Center. This work is supported by DOE contract DE-FG0291ER40688,
Tasks A and D.

%% missing bibliographystyle here?
\bibliography{emirdag}
\end{document}